\documentclass[useAMS,usenatbib]{mn2e}

\def\Msol{\hbox{M$_\odot$}}

\def\kms{\hbox{km$\,$s$^{-1}$}}

\def\Apix{\hbox{\AA$\,$pix$^{-1}$}}
\def\one{\,{\sc i}}             
\def\two{\,{\sc ii}}
\def\three{\,{\sc iii}}

\newcommand\fsec{\hbox{$.\!\!^{\rm s}$}}
\def\farcmin{\hbox{$.\mkern-4mu^\prime$}}

\usepackage{amsmath}
\usepackage{amssymb}
\usepackage{mathrsfs}
\usepackage{color}
\usepackage{soul} 
\usepackage{upgreek}
\usepackage{graphicx}
\usepackage{overpic}

\defcitealias{westm09a}{Paper~I}
\defcitealias{westm09b}{Paper~II}
\defcitealias{mckeith95}{McK95}
\defcitealias{mckeith93}{McK93}

\title[Kinematics of the multi-phase interstellar medium in M82]{Spatially resolved kinematics of the multi-phase interstellar medium in the inner disk of M82}
\author[M.S.\ Westmoquette et al.] {M.\ S.\ Westmoquette$^1$\thanks{E-mail: mwestmoq@eso.org}, L.\ J.\ Smith$^2$, J.\ S.\ Gallagher III$^3$, and F.\ Walter$^4$ \\
$^1$European Southern Observatory, Karl-Schwarzschild-Str. 2, 85748 Garching bei M\"{u}nchen, Germany\\
$^2$Space Telescope Science Institute and European Space Agency, 3700 San Martin Drive, Baltimore, MD 21218, USA\\
$^3$Department of Astronomy, University of Wisconsin-Madison, 5534 Sterling, 475 North Charter St., Madison WI 53706, USA\\
$^4$Max-Planck-Institut f\"{u}r Astronomie, K\"{o}nigstuhl 17, D-69117 Heidelberg, Germany \\
}
\date{}
\pagerange{\pageref{firstpage}--\pageref{lastpage}}
\pubyear{2012}
\begin{document}
\maketitle
\label{firstpage}
\begin{abstract}
We present spatially resolved kinematics of the interstellar Na\one\ D $\lambda\lambda$5890,5896 doublet absorption and $^{12}$CO~(1$\rightarrow$0) emission across the inner $\sim$2\,$\times$\,1~kpc of the disk of M82. These data were obtained with the DensePak IFU on the WIYN telescope and the Caltech Owens Valley Radio Observatory (OVRO) millimetre array. By measuring the Na\one\ and CO (and H$\alpha$ kinematics from a previous study) at the same spatial resolution, and employing the same line fitting method, we have been able to make meaningful comparisons between the ionized, neutral and molecular gas phases. We detect a component of the Na\one\ line throughout the inner disk with velocities that are forbidden by the known galactic rotation. We interpret this as originating in counter-rotating or perhaps inflowing material. In the southern plume, we find clear evidence of entrained CO gas with kinematics consistent with that of H$\alpha$. On the northern side, the CO kinematics appear to trace more static clouds in the inner halo that could be pre-existing or tidal in origin. We find no evidence that Na\one\ absorption is kinematically associated with the outflow. We conclude that a combination of lack of velocity resolution and confusion of due to the high inclination of the system is acting to prevent detection. Thus, in the search for neutral outflows from galaxies, the signature high velocity components may easily be missed in observations at low spectral resolution and/or sensitivity, and particularly so in highly inclined systems.
\end{abstract}

\begin{keywords} galaxies: individual (M82) -- galaxies: starburst -- galaxies: ISM -- ISM: kinematics and dynamics.
\end{keywords}

\section{Introduction}\label{intro}

M82 is the archetype nearby \citep[3.6~Mpc, $1'' = 17.5$~pc;][]{freedman94} starburst galaxy \citep{oconnell78, oconnell95}. The current ($\sim$10~Myr) starburst activity is concentrated in a $\sim$500~pc ($\sim$30$''$) region centred on the nucleus. From \textit{Hubble Space Telescope} (\textit{HST}) imaging, the starburst is known to consist of many young massive star clusters \citep{oconnell95, mccrady03}, and it is the energy from these clusters that drives the famous H$\alpha$- and X-ray-bright, kpc-scale, superwind \citep{shopbell98, ohyama02, stevens03a, strickland07, westm07c, westm09a}. 

The bipolar outflow \citep{shopbell98, engelbracht06, mutchler07} can be observed easily due to the galaxy's almost edge-on inclination \citep[$i \sim 80^{\circ}$;][]{lynds63, mckeith95}. \citet[][hereafter \citetalias{mckeith95}]{mckeith95} presented deep optical and near-infrared long-slit observations of emission and absorption lines along M82's minor axis, and found clear line splitting in their emission lines. This was interpreted as originating from the front and back walls of a cone-shaped structure, inclined such that the southern cone is directed toward the observer. 

Radial velocities along the major axis have been studied many times in the optical/near-IR (e.g.\ \citealt{burbidge64,  mckeith93, sofue98}; \citealt{shopbell98}; \citealt{castles91, westm07c}), IR \citep[e.g.][]{larkin94, achtermann95} and radio \citep[e.g.][]{wills00, walter02, r-r04}. Overall, these studies have found an inner $\sim$$5''$ region with a steep velocity gradient, and an outer region $>$$\pm$$25''$ of constant (or slowly increasing) velocities. These gradients are centred on the systemic velocity of the galaxy, $v_{\rm sys}=+200$~\kms. 

The steep inner region is attributed to the $x_{2}$-orbits of a $\sim$1~kpc long stellar bar, known from near-infrared and H\one\ studies \citep{larkin94, achtermann95, wills00}. The interaction between the outer $x_{1}$ and inner $x_{2}$ orbit families has resulted in the build up of a torus of molecular and ionized gas and dust at a radius of $\sim$250~pc \citep{achtermann95, weis01}.

\citet[][hereafter \citetalias{westm09a}]{westm09a} examined the ionized gaseous and stellar dynamics of M82's disk and inner wind via DensePak and Gemini/GMOS spatially resolved optical integral field unit (IFU) spectroscopy. 
We found clear evidence for the wind emanating from the entire starburst zone. We also found a broad component in the ionized emission lines that we interpreted as tracing turbulent mixing interfaces between the hot wind and photoionized gas phases. In \citet[][hereafter \citetalias{westm09b}]{westm09b} we examined the nebular properties in more detail, and described the discovery of a distinct channel $>$100~pc in length identified within the inner wind, traced by the ionized gas and characterised by its dynamics and density morphology.

The central starburst is believed to be fuelled by the large amount of cold molecular gas present in the disk \citep[$M_{\rm tot} = 2.3\times 10^{8}$~\Msol;][]{weis01, garcia-burillo02}. \citet{walter02}, however, discovered significant amounts of molecular material ($1.3\times 10^{9}$~\Msol) outside the central 1~kpc disk. Some streamers were found to be well correlated with optical absorption and H\one\ features, implying that the molecular gas has been partly disrupted by the tidal interaction with M81. However, CO line splitting was also detected for the first time, implying that there is also significant molecular material entrained in the wind ($\sim$$3\times 10^{8}$~\Msol).

While much work has been done on the distribution of ionized gas in nearby starburst outflows \citep[e.g.][]{marlowe95, martin98, westm07a, sharp10, westm11}, in comparison very little has been done (particularly on spatially extended scales) on the neutral phases in these same nearby examples. Cool neutral gas is most easily probed through absorption lines. Due to their advantage in (1) being able to be seen against a bright continuum source (continuum surface brightnesses are typically much larger than that of the emission lines), and (2) mitigating the ambiguity in the radial flow direction, absorption lines have been used extensively to probe outflows in more distant sources \citep[$z\gtrsim1$, e.g.][]{shapley03, fox07, weiner09, rubin10, kornei12, martin12}.

Since, in these samples, the absorption kinematics are often treated as direct probes of the outflow speed, it is important to examine whether this really is the case in nearby systems. \citet{schwartz04} presented Keck/HIRES Na\one\ D absorption line observations of a sample of nearby dwarf starbursts and found that the kinematics of Na\one\ D and H$\alpha$ were generally very different. \citet{rupke05a} compared the ionized emission and Na\one\ D absorption line profiles for a sample of 17 starburst-dominated ULIRGs, and also found a complicated situation. Only in some cases blueshifted wings were detected in both phases. In half of those, the ionized gas had a higher maximum velocity, while in the other half they were equivalent. The situation, therefore, is far from clear.


In this paper we aim to determine the kinematics of both the cool neutral and molecular media across the entire inner $\sim$$2\times1$~kpc of M82's disk using spatially resolved (3$''$ resolution) observations of the interstellar Na\one\ D $\lambda\lambda$5890,5896 doublet absorption and the $^{12}$CO (1$\rightarrow$0) (2.6~mm, 115~GHz) emission line. By comparing these results to those of the ionized gas \citep[made at the same resolution;][]{westm09a}, we can make a direct and meaningful comparison between the three gas phases.



\section{Observations} \label{sect:obs}

\subsection{DensePak observations} \label{sect:DP_obs}
On 14th April 2001, we observed four fields covering the stellar disk of M82 with the DensePak instrument \citep{barden98}, a small fibre-fed integral field array attached to the Nasmyth focus of the WIYN (Wisconsin, Indiana, Yale and NOAO) 3.5-m telescope. It has 91 circular fibres, each with a diameter of 300\,$\micron$ or $3''$ on the sky; the fibre-to-fibre spacing is 400\,\micron\ making the overall dimensions of the array $30\times 45$~arcsec. The format of the array, including the gaps between the fibres, is represented in the maps presented below. DensePak's fibre bundle is reformatted into a pseudo-slit to feed the WIYN bench-mounted echelle spectrograph.

\begin{figure*}
\centering
\includegraphics[width=13cm]{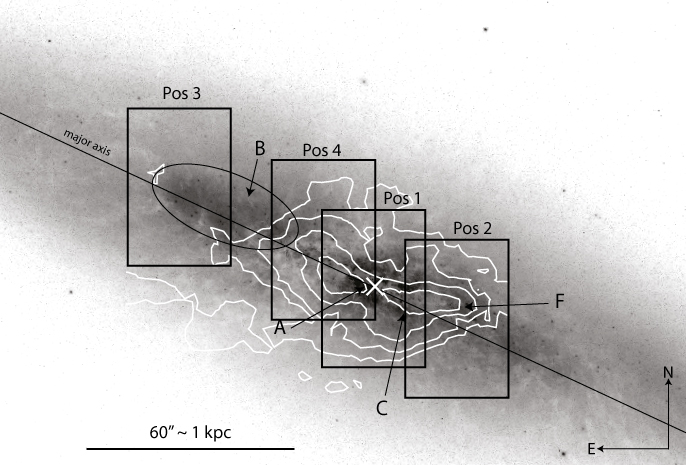}
\caption{\textit{HST}/ACS F814W (I-band) image with the DensePak IFU footprints overlaid, together with arrows labelling certain well-known regions within the central starburst. The 2.2~\micron\ nucleus \citep[][$\alpha= 09^{\rm h}51^{\rm m}43\fsec4$, $\delta = 69^{\circ}55'00''$, B1950]{lester90} is marked by a white cross. Contours show the integrated $^{12}$CO flux \citep{walter02}.}
\label{fig:HSTfinder}
\end{figure*}

The spectrograph was set up to give a dispersion of 0.46~\Apix\ over the wavelength range of 5820--6755\AA. This allowed us access to a number of optical nebular lines (analysed in \citetalias{westm09a} and \citetalias{westm09b}) and the Na\one\ D $\lambda$$\lambda$5890,5896 doublet which we examine here.  Na\one\ D absorption originates from warm neutral gas in the interstellar medium and stellar atmospheres. We refer the reader to \citetalias{westm09a} for further details regarding the IFU position coordinates, exposure times and data reduction methodology. Fig.~\ref{fig:HSTfinder} shows the position of the DensePak fields overlaid on an \textit{HST}/ACS F814W (I-band) image. The instrumental resolution is $44.7\pm 3.1$~\kms\ \citepalias{westm09a}.

\subsubsection{Decomposing the Na\one\ line profiles} \label{sect:line_profiles}

Na\one\ absorption is detected in every fibre of the DensePak fields, and the high S/N and spectral resolution of our data have allowed us to quantify the line profile shapes to a high degree of accuracy. Because of the coincidence of telluric Na\one\ emission (mostly from the street lights in Tucson) with the interstellar absorption lines, accurately measuring the absorption profile shape is highly dependent on the sky subtraction quality. Thus, to avoid the introduction of artefacts that are always associated with attempting to remove the sky lines, we here perform our analysis on the original spectra (i.e.\ with no sky subtraction).

For each spectrum we simultaneously modelled the Na\one\ doublet absorption together with the He\one~$\lambda$5876 nebular emission line and the three telluric emission lines at 5888.05, 5889.95, and 5895.92\,\AA\ (the latter two being the Na\one\ doublet) using using the \textsc{idl} $\chi^2$ fitting package \textsc{pan} \citep[Peak ANalysis][]{dimeo, westm07a}. Milky Way absorption was filled in by the telluric emission so was not included in the fit. In order to limit the number of free parameters as much as possible, we fitted the Na\one\ doublet (both emission and absorption) as a single model consisting of two Gaussians with wavelength separations equal to the lab-measured difference, and equal line widths specified to be greater than the instrumental resolution. The fluxes of the two lines of the doublet were left unconstrained. We fitted each of the Na\one\ absorption profiles with a single and double component initial guess. Multi-component fits were run several times with different initial guess configurations (widths and wavelengths) in order to account for the varied profile shapes, and the one with the lowest $\chi^2$ fit statistic was kept. However, we note that the $\chi^2$ minimisation routine employed by \textsc{pan} is very robust with respect to the initial guess parameters.

To determine whether the Na\one\ doublet needed one or two components to fit its shape, we used a combination of visual inspection, the $\chi^{2}$, and the fit residuals. In the majority of spectra, two components were needed. In these cases, we named the stronger absorption component C1 and the weaker C2. Fig.~\ref{fig:eg_fits} shows some example spectra in the Na\one\ region, showing simultaneous fits to the He\one~$\lambda$5876 emission line, the Na\one\ absorption lines and the three telluric emission lines (note that He\one\ is not detected in two of these examples). These spectra were extracted from the spaxels labelled with the corresponding letters in Fig.~\ref{fig:vel_NaI} (see below).

Due to the constraints imposed on the telluric line fits, formal errors on the Na\one\ fit results are not straightforward to calculate. We estimate that uncertainties in the radial velocities of the Na\one\ results are 5--10~\kms\ for high S/N spaxels and remain $\le$30~\kms\ for the majority of spectra. FWHM errors range between $\le$10~\kms\ for high S/N spectra and remain below $\sim$50~\kms\ for most spaxels. Uncertainties in the absorption equivalent widths (W$_{\rm eq}$) of the components range between 0.1~\AA\ for the strongest absorption to $\sim$0.5~\AA\ for the weaker components. In some cases, these high uncertainties arise from the $\lambda$5896 (redder) Na\one\ emission line falling on the red wing of the bluer ($\lambda$5890) absorption line.

Fig.~\ref{fig:ew_NaI} shows the Na\one\ W$_{\rm eq}$ maps for components C1 and C2. The W$_{\rm eq}$ of Na\one\ C2 is a factor of $\sim$3--5 times smaller than C1. As a check on the physicality of our final results, we measured the ratio of the equivalent widths of the two Na\one\ lines for each fitted component, since a ratio of W$_{\rm eq,5890}$/W$_{\rm eq,5896}$ = 1 represents the optically thick limit, whereas a ratio of 2 represents the optically thin limit. As expected, in almost every case the ratio is $\sim$1. However, in spectra where the uncertainties in the W$_{\rm eq}$ of the weaker component are high, this ratio is often $<$1 (Fig.~\ref{fig:eg_fits}d is an obvious case). While this reflects the inherent difficulty of fitting such a complex and blended set of lines, we believe that, although the equivalent widths of the weaker components may be uncertain, the kinematics are accurate.


\subsection{OVRO CO map} \label{sect:CO_obs}
Observations of M82 in $^{12}$CO (1$\rightarrow$0) (2.6\,mm, 115\,GHz) were made with the Caltech Owens Valley Radio Observatory (OVRO) millimetre array over a a $2\farcmin5$ $\times$ $3\farcmin5$ field at a spatial resolution of $3\farcs6$, and were presented in \citet{walter02}. More details on the setup and data reduction methodology are given in that paper. Here we use these data to compare to our ionized and neutral phase measurements.

First we cropped the cube to the region coincident with our DensePak data, then binned it spatially to a resolution of $3''$ to match that of the DensePak fibres. We then masked any spaxels with integrated fluxes below 15~Jy~beam$^{-1}$ (equating to a line S/N below $\sim$3) in order to ensure that every spatial element in our final cube contained a significant CO line detection.

In order to produce measurements for the CO data that are comparable to our ionized and neutral gas measurements, we fitted the CO line using \textsc{pan} in much the same way as we have done for the ionized emission lines (\citetalias{westm09a}) and Na\one\ absorption lines. We found we required two Gaussian emission components in the majority of spaxels in the cropped and masked cube. Whenever two components were needed, we assigned the component whose radial velocity was closest to $v_{\rm sys}$ (+200~\kms) to C1 and the higher velocity component to C2. We found that this resulted in a more coherent velocity map than if we had followed the convention adopted for Na\one\ (where we assigned the stronger component to C1 and the weaker to C2 regardless of velocity). \footnote{In Na\one, the stronger component dominates the absorption profile and is the one that mainly follows galactic rotation (see Sect~\ref{sect:radvel_maps}), whereas in CO the relative strength of the two components varies much more, meaning that a velocity-based assignment is more appropriate.} Fig.~\ref{fig:vel_CO} shows the resulting line centroid velocity maps for the two components, and Fig.~\ref{fig:eg_fitsCO} shows some example fitted CO line profiles extracted from selected regions labelled with the corresponding letters in Fig.~\ref{fig:vel_CO}.

\section{Results} \label{sect:results}

\subsection{Radial velocity maps}\label{sect:radvel_maps}
In Fig.~\ref{fig:vel_NaI} we present the Na\one\ line centroid radial velocity maps for the two identified line components. Recall that for Na\one, the fainter component of any two-component fits was assigned to C2. The signature of galactic rotation is seen in the C1 component, with approaching velocities in the west and receding velocities in the east, although the velocity gradient is much shallower than that of the stellar or ionized emission lines (\citetalias{westm09a}, and see below). In \citetalias{westm09a} we confirmed a 12$^{\circ}$ offset to the position angle of the ionized gas rotation axis, compared to both the stellar rotation axis and photometric major axis, that had first been seen by \citet{r-r04}. The line connecting the maxima and minima of this offset rotation axis is shown with a dashed line in the Na\one\ C1 map (Fig.~\ref{fig:vel_NaI}). Despite the shallower velocity gradient compared to that of the ionized gas (see below), it is clear that the rotation seen in Na\one\ C1 is consistent with same offset axis (blueshifted velocities in the west peak below the major axis, whereas redshifted velocities in the east peak above the major axis).

\begin{figure*}
\includegraphics[width=\textwidth]{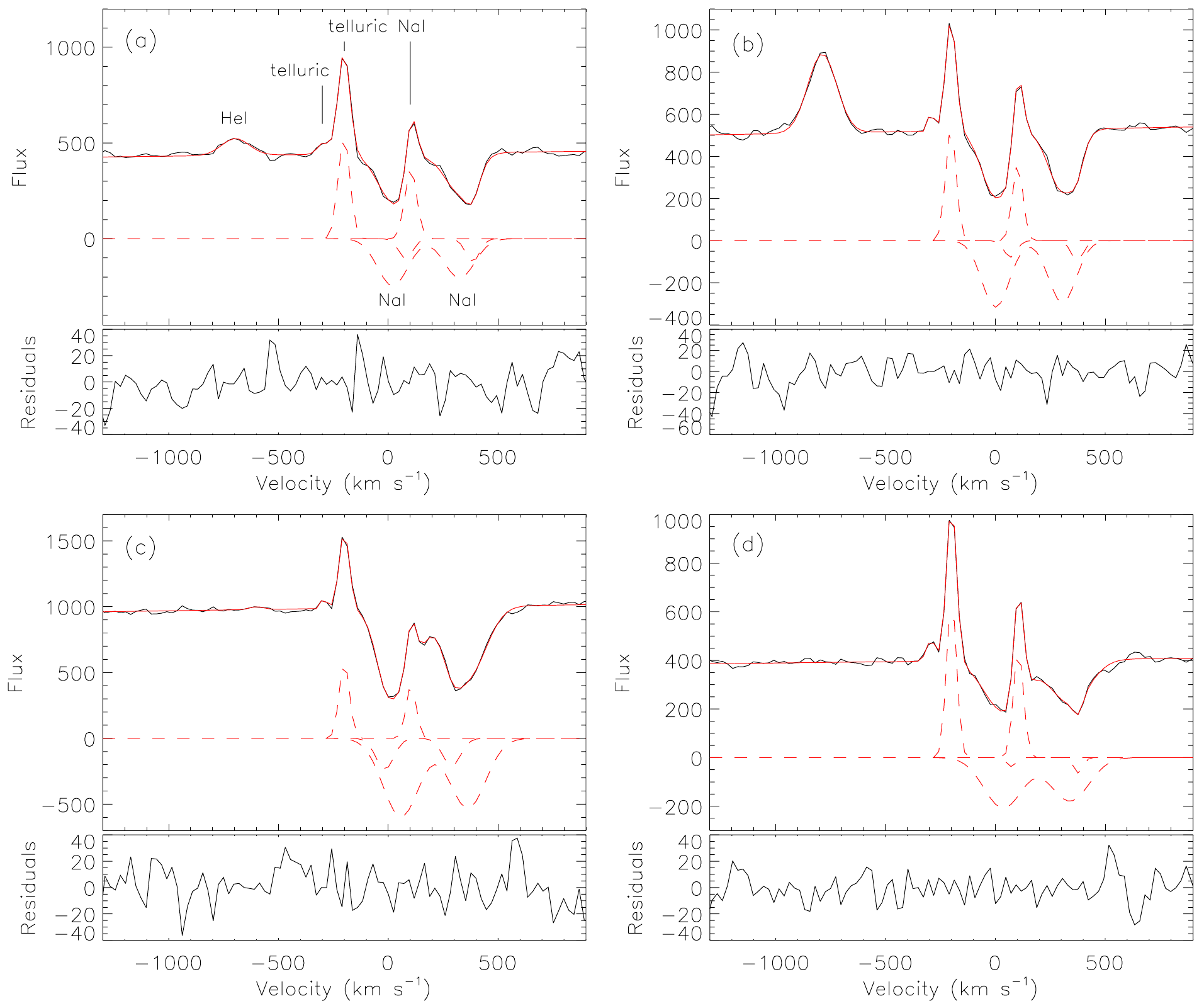}
\caption{Examples of fits to the Na\one\ lines, including the He\one~5876\AA\ and telluric emission lines as labelled in panel (a). Zero velocity corresponds to the rest wavelength of Na\one\ $\lambda$5890. The $y$-axes are in arbitrary but relative flux units. The location from which these spectra originate in the IFU fields are indicated with the corresponding letters in the left panel of Fig.~\ref{fig:vel_NaI}.}
\label{fig:eg_fits}
\end{figure*}

\begin{figure*}
\includegraphics[width=\textwidth]{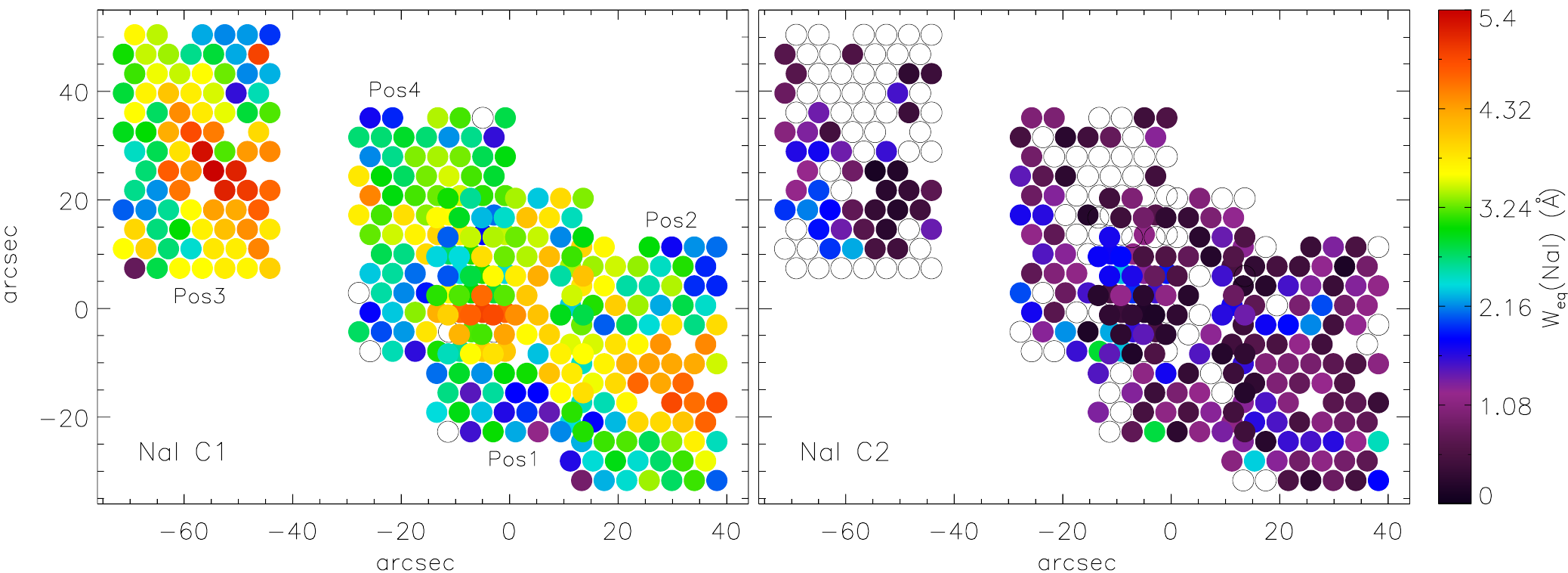}
\caption{Map of the equivalent width of Na\one\ absorption for the two line components. The (0,0) origin is centred on the 2.2~\micron\ nucleus.}
\label{fig:ew_NaI}
\end{figure*}

\begin{figure*}
\includegraphics[width=\textwidth]{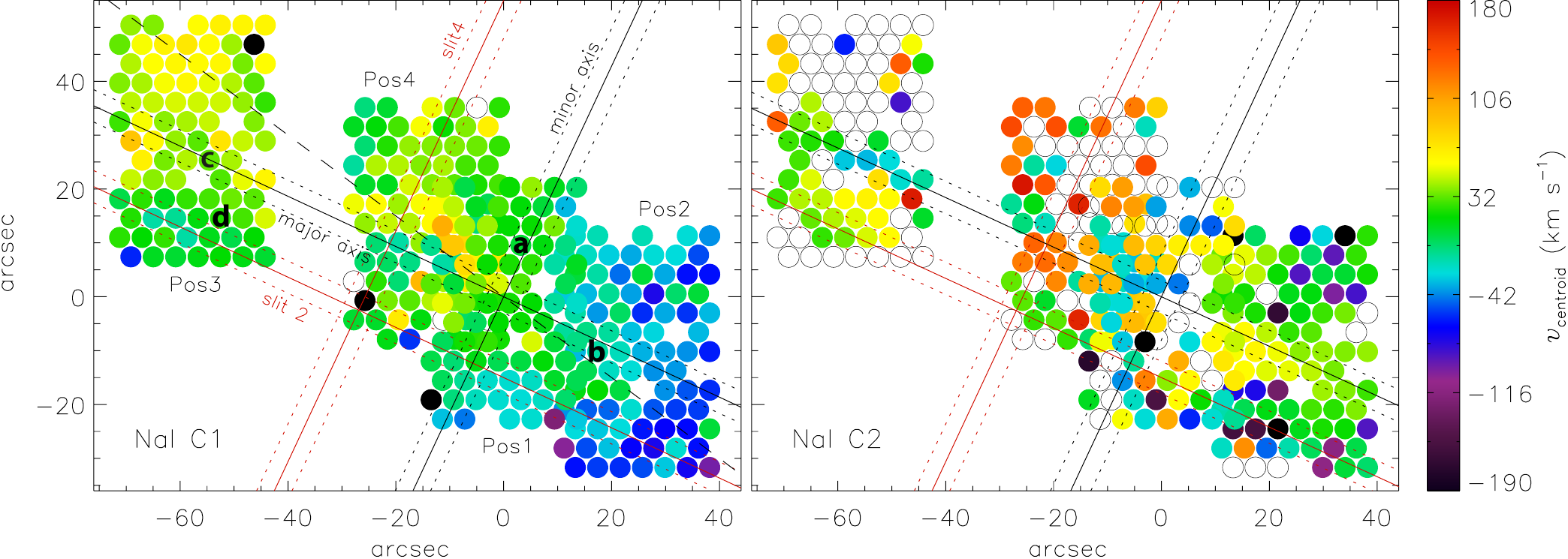}
\caption{Na\one\ line centroid velocity maps for components C1 and C2 relative to $v_{\rm sys}$ (=+200~\kms). The (0,0) origin is centred on the nucleus. Letters in the left panel show the location of spectra shown in Fig.~\ref{fig:eg_fits}. The galaxy major and minor axes are marked with solid black lines. Pseudoslits along these axes, together with two additional pseudoslits (labelled slit 2 and 4) offset by 15$''$ to the south of the major axis and 20$''$ to the east of the minor axis (drawn in red), were used to make the position-velocity diagrams of Figs.~\ref{fig:dp_major_axis} and \ref{fig:dp_minor_axis}. The pseudoslit widths are indicated by the dotted lines. The dashed line in the left panel represents the PA of the gaseous (H$\alpha$, [S\three], P9) rotation axis (offset by $\sim$12$^{\circ}$; determined from the emission line maps in \citetalias{westm09a}).}
\label{fig:vel_NaI}
\end{figure*}

\begin{figure*}
\includegraphics[width=\textwidth]{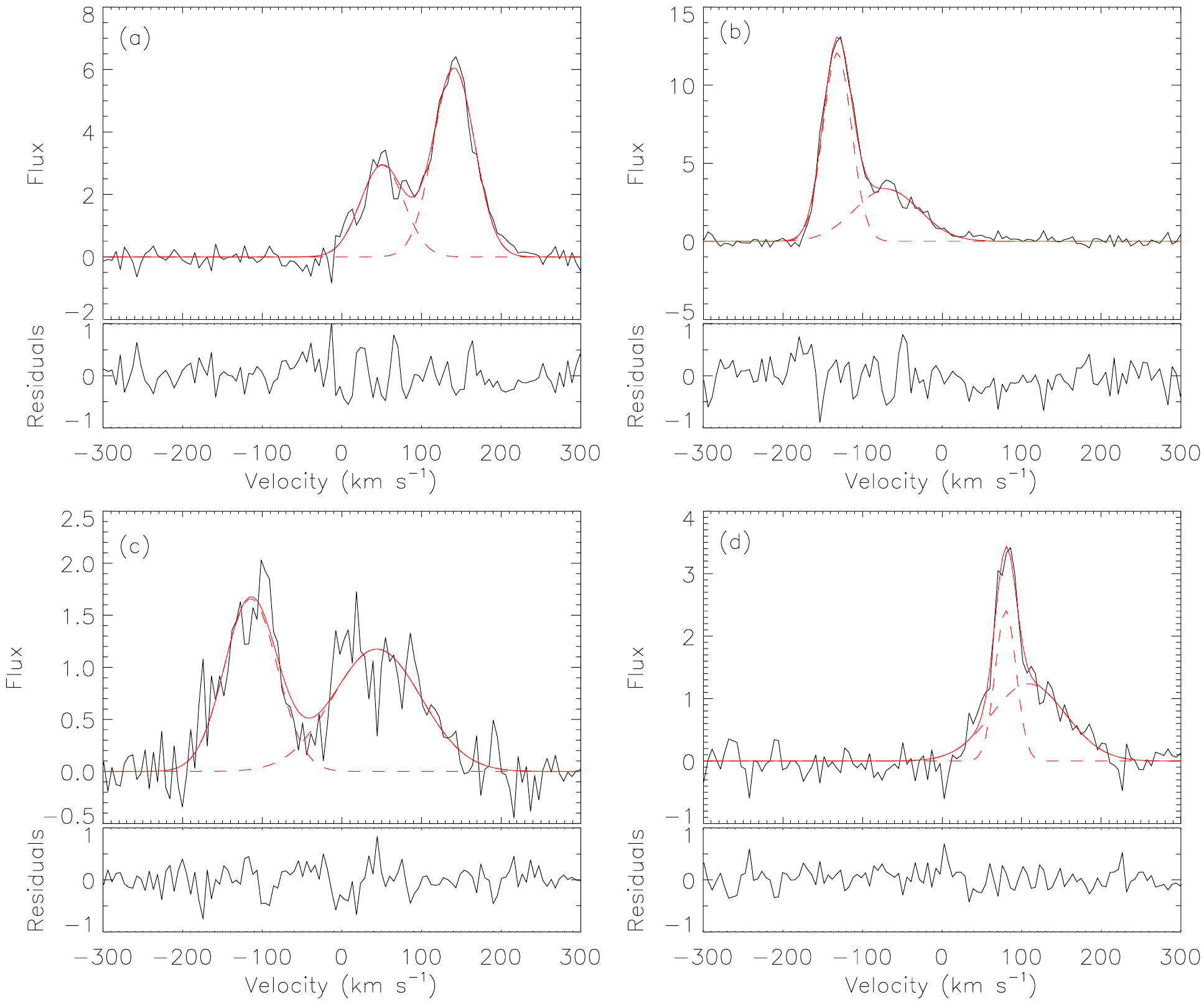}
\caption{Examples of fits to the CO line from selected locations as indicated with the corresponding letters in the left panel of Fig.~\ref{fig:vel_CO}. The fluxes are in units of Jy~beam$^{-1}$.}
\label{fig:eg_fitsCO}
\end{figure*}

\begin{figure*}
\includegraphics[width=\textwidth]{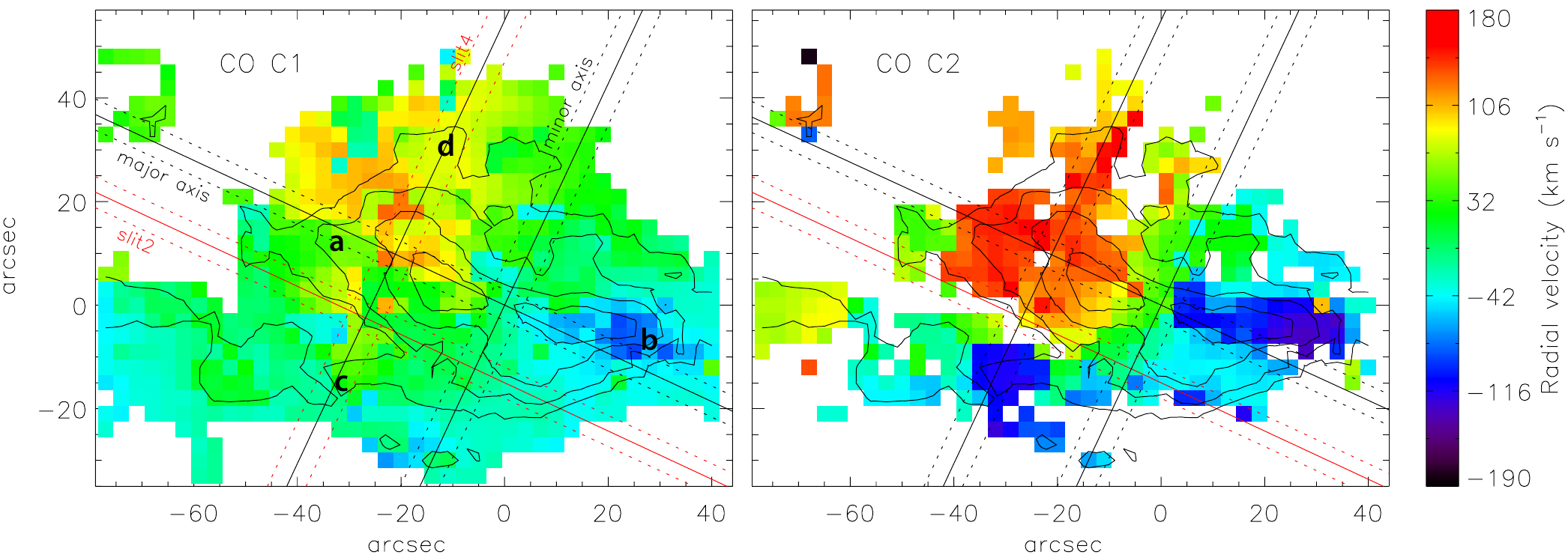}
\caption{$^{12}$CO (1$\rightarrow$0) centroid velocity maps for components C1 and C2, relative to $v_{\rm sys}=200$~\kms. The overlaid contours represent the integrated CO flux, and clearly show the two molecular lobes.
Letters in the left panel show the location from which the spectra shown in Fig.~\ref{fig:eg_fitsCO} were extracted. The same pseudoslits as shown in Fig.~\ref{fig:vel_NaI} (used for the p-v diagrams of Figs.~\ref{fig:dp_major_axis} and \ref{fig:dp_minor_axis}) are shown.}
\label{fig:vel_CO}
\end{figure*}

The dynamics of the fainter Na\one\ component (C2) are much more complex. To the west of the minor axis, C2 is redshifted compared to C1 (e.g.\ Fig.~\ref{fig:eg_fits}b). To the east, some spaxels contain a blueshifted C2 component whereas others show a redshifted one. In position 3, a handful of spaxels along the midplane exhibit a blueshifted C2 component (e.g.\ Fig.~\ref{fig:eg_fits}c), although most C2 velocities in the south of the position are redshifted with respect to C1 (e.g.\ Fig.~\ref{fig:eg_fits}d).

In Fig.~\ref{fig:vel_CO} we present the $^{12}$CO emission line radial velocity maps. Two line components are detected throughout the central 80$''$$\times$60$''$ (1400$\times$1050~pc). Unlike the flux-based component assignments we used for Na\one, for CO we assigned the components based on velocity, with the one closest to $v_{\rm sys}$ as C1 and the higher velocity component as C2. The rotation of the central CO torus \citep{shen95} can clearly be seen, with the redshifted receding side in C1 and the approaching blueshifted side in C2. 

The differences between the Na\one\ and $^{12}$CO maps and the H$\alpha$ maps from \citetalias{westm09a} clearly indicates that the gas kinematics within this central starburst region are very complex. To investigate these results in more detail we define a number of pseudoslits along and parallel to the major and minor axes from which we extract the line component velocities. These are discussed in the following sections.

\subsection{Major axis position-velocity graph} \label{sect:DP_major}
\begin{figure*}
\centering
\includegraphics[width=17cm]{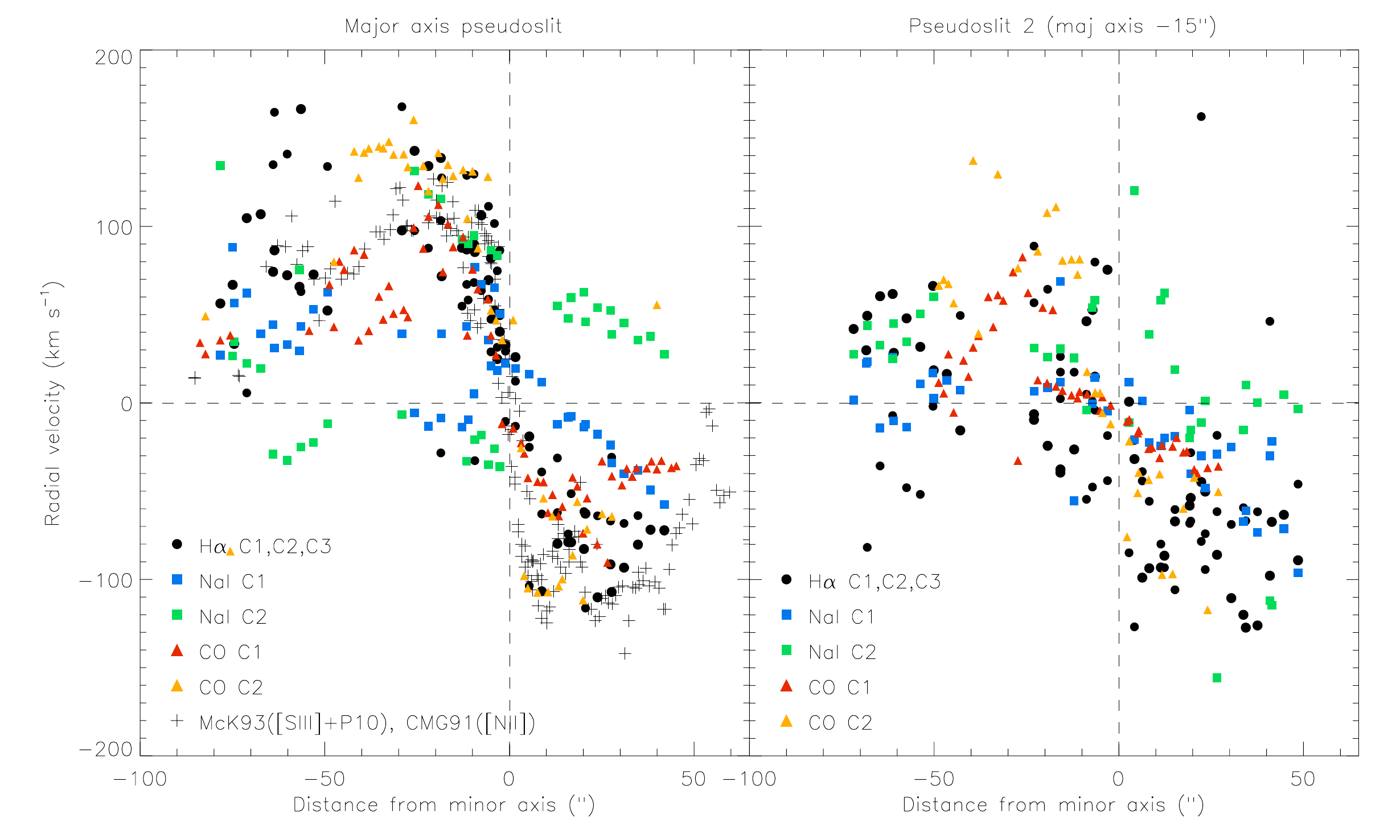}
\caption{Major axis position-velocity diagrams for Na\one, CO and H$\alpha$ \citepalias{westm09a}, extracted from the corresponding $7''$ wide pseudoslits shown in Figs~\ref{fig:vel_NaI} and \ref{fig:vel_CO}. Velocities are shown with respect to $v_{\rm sys}$. 
The optical and near-IR ionized gas velocities measured along the major axis by \citet{castles91} and \citet{mckeith93} are plotted with + symbols.}
\label{fig:dp_major_axis}
\end{figure*}

Position-velocity plots extracted from two $7''$-wide pseudoslits positioned along the galaxy major axis and offset 15$''$ to the south are shown in Fig.~\ref{fig:dp_major_axis} (left and right panels respectively). The location of these pseudoslits are shown in both Fig.~\ref{fig:vel_NaI} and \ref{fig:vel_CO}. We plot our new Na\one\ and $^{12}$CO measurements together with the H$\alpha$ measurements from \citetalias{westm09a}, the [N\two]$\lambda$6583 velocities from \citet{castles91}, and the [S\three], P10, and Ca\two\ velocities from \citet[][+ symbols]{mckeith93}. 

A close inspection of these two plots reveals a number of points of interest. Between $-20$ to +5$''$ on the major axis (the eastern side of the inner $x_{2}$ bar region) the H$\alpha$, Na\one\ and CO measurements are mostly in good agreement, indicating that we are seeing molecular, neutral and ionized gas all following the bar orbits. However, on the western side (mostly between $-10$ and +40$''$), the Na\one\ C1 follows a distinctly different, shallower velocity gradient. In pseudoslit 2, the Na\one\ C1 gradient is flatter still, and between $-20$ to +20$''$ is followed by the CO. These shallower gradients imply that this neutral and molecular gas must be located nearer to us along the line-of-sight (i.e.\ at larger galactocentric radii) where the radial component of the orbital velocity vector is less.

Between $-65$ and 0$''$ in the major axis pseudoslit we find a consistent area of blueshifted (10--60~\kms) Na\one\ C2 points, and between +10 and +45$''$ we find a cluster of redshifted Na\one\ C2 points (30--70~\kms). Example line profiles and fits for these regions are shown in Fig.~\ref{fig:eg_fits}b and c. In the velocity map (Fig.~\ref{fig:vel_NaI}) we see that these redshifted C2 velocities are present throughout the western section of the disk (as mentioned above). Together, these points appear to trace a component of neutral material with very different kinematics to the rest of the gas.


\subsubsection{Minor axis position-velocity graph} \label{sect:DP_minor}

Fig.~\ref{fig:dp_minor_axis} shows position-velocity plots from two $7''$-wide pseudo-slits oriented along the minor axis (left panel) and a parallel line 20$''$ to the east (pseudoslit 4; right panel). The location of these pseudoslits are also indicated in Figs.~\ref{fig:vel_NaI} and \ref{fig:vel_CO}. As for the major axis plots, we include our new Na\one\ and CO measurements together with the H$\alpha$ emission line velocities from \citetalias{westm09a}. We also include the H$\alpha$, [N\two]$\lambda$6583 and [S\three]$\lambda$9532 data from \citet{mckeith95}. 

From $+10''$ northwards along the minor axis (left panel) the CO stays at a very constant velocity and is considerably redshifted from the ionized gas. In the north of pseudoslit 4, CO C2 traces H$\alpha$, whereas CO C1 stays at a very constant velocity (from $\sim$15--35$''$), remaining on the blue side of C2. The position of these points in the two p-v diagrams suggest that they may be tracing the far-side of the northern wind outflow cone, although their very constant velocities are suggestive of more quiescent gas clouds in the outer disk. We discuss this below in Section~\ref{sect:molecular}.


From $-10''$ southwards the Na\one\ and CO velocities agree well with ionized gas velocities, suggesting that here both the Na\one\ and CO trace the far-side of the wind cone. At the equivalent offsets in pseudoslit 4, the CO emission is composed of two distinct velocity components. The redshifted component (C2, becoming C1) accelerates at a $\sim$constant rate from $-5''$ out to $>$$-35''$ ($>$600~pc), consistent with a number of H$\alpha$ points between 0 and $-15''$. Between 0 and $-15''$, the bluer component (C1; here blueshifted from C2 by $\gtrsim$100~\kms), traces the kinematics of Na\one\ C1 and H$\alpha$ , then jumps to $\sim$$-110$~\kms\ from $-20''$ southwards (and assigned to C2). These split line components are consistent with tracing the wind outflow, as was noted by \citet{walter02}.

\section{Discussion} \label{sect:disc}

M82 is a complex system that is not easily interpreted. The kinematics of any one gas phase in the central starburst region are complex down to the resolution limit of any observation made to date. Understanding the relationship between the different gas phases is therefore challenging, particularly when the sensitivity, spatial resolution and/or observation method of the data are not matched. In this study we have attempted to compare the neutral Na\one\ absorption and $^{12}$CO emission kinematics to the ionized (H$\alpha$) emission line kinematics at the same spatial resolution, employing the same line fitting method for each. In the following we discuss some of the salient points from our results in context with the literature and the M82 system as a whole.


\subsection{Neutral gas in the wind} \label{sect:neutral}


We find no evidence of Na\one\ absorption that is kinematically associated with warm ($\sim$$10^4$~K) ionized gas outflow from our DensePak data. This could be due to (1) there being no cool neutral gas in the outflow, (2) a lack of velocity resolution meaning that weak blueshifted outflow components are unresolved or washed out in our spectra, or (3) that the high inclination of the system means that the absorption line is confused with multiple line-of-sight components. Regarding point (1), it would be rather surprising that the outflow does not entrain any cool neutral material, given that we know there is ionized and molecular gas in the flow.

We already know from Na\one\ radial velocity measurements along sight-lines towards star clusters \citep{konstantopoulos09} that the Na\one\ is decoupled from the stars, clusters, and ionized gas. Although the clusters measured by \citet{konstantopoulos09} were not strictly along the major axis, the overall Na\one\ radial velocity distribution within the central $\sim$2~kpc of the disk is consistent with a flat gradient, suggesting that the absorbing neutral material is located further out in the disk (where the radial component of its orbital velocity is smaller). The high wind speeds in M82 should rapidly fragment clouds, so perhaps it would not be a surprise for Na\one\ absorption from entrained clouds to be patchy \citep[this effect can be seen in simulations such as][]{cooper09}. Having said that, strong signatures of entrained Na\one\ absorbers are found in many other starburst systems \citep[e.g.][]{rupke05a, chen10}.

On points (2) and (3), the velocity resolution of our DensePak spectra ($\sim$45~\kms) certainly does prevent us from being able to see the finer details of the Na\one\ absorption profile, known from previous higher resolution observations \citep{smith_gall01, schwartz04}. In Fig.~\ref{fig:m82-f_comp} we compare the Na\one\ line profile of M82-F, a super star cluster to the west of the nucleus (see Fig.~\ref{fig:HSTfinder}), from our DensePak data to that measured by \citet{smith_gall01} using the Utrecht Echelle Spectrograph (UES) on the William Herschel Telescope (WHT) at a resolution of 8~\kms\ (R=45000). Whereas five sharp absorption components can be identified in the UES spectrum, at the lower resolution of our DensePak spectrum only two (broader) components can be resolved. 

\citet{schwartz04} observed a $\sim$4$''$$\times$$1\farcs5$ sight-line towards the centre of M82 clump A (centred at coordinates $09^{\rm h}\,55^{\rm m}\,52.6^{\rm s}$, $+69^\circ\,40'\,47\farcs0$, J2000; priv.\ comm.) at 11~\kms\ (R=30000) resolution with Keck/HIRES. They also found an extremely complex absorption system best fitted with five separate components, the two strongest being blueshifted by $-90$ and $-35$~\kms.

These higher resolution observations show that the neutral gas kinematics in M82 are extremely difficult to disentangle, and great care must be taken. The presence of these narrow blueshifted components (in both the UES and HIRES spectra) suggests that there is indeed neutral gas entrained in the outflow which cannot be seen at the resolution of our DensePak spectra. Furthermore, the presence of so many components over a large range in velocities belies the fact that, due to the high inclination of the galaxy, these absorption profiles are likely to be quite confused by material along the line-of-sight, particularly in these central regions of the disk.

\subsection{Molecular gas in the wind} \label{sect:molecular}

\begin{figure*}
\centering
\includegraphics[width=13cm]{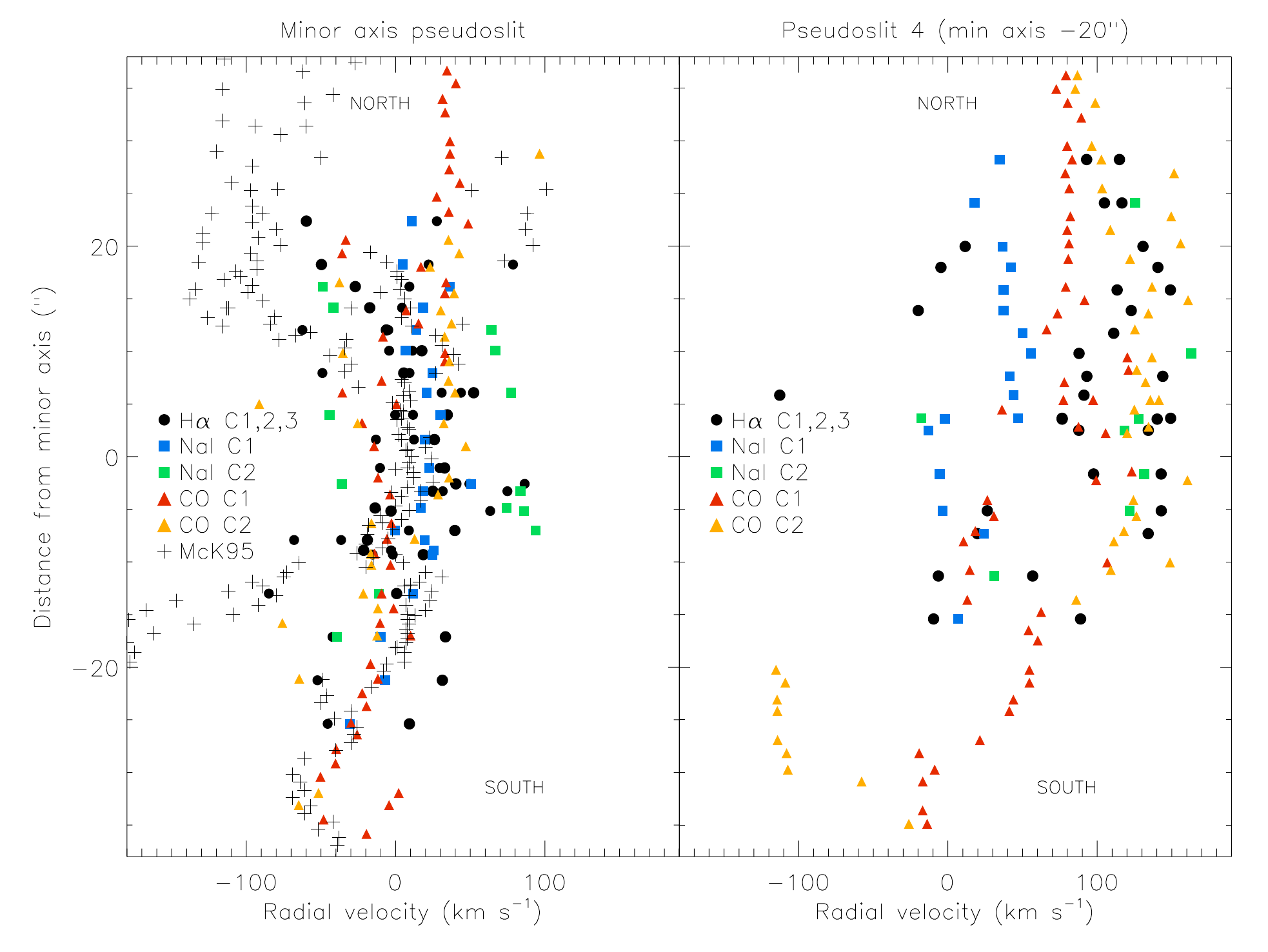}
\caption{Minor axis position-velocity diagrams for Na\one, CO and H$\alpha$ \citepalias{westm09a}, extracted from the corresponding $7''$ wide pseudoslits shown in Figs.~\ref{fig:vel_NaI} and \ref{fig:vel_CO}. Velocities are shown with respect to $v_{\rm sys}$. 
The optical and near-IR ionized gas velocities measured along the major axis by \citet{mckeith95} are plotted with + symbols. The letter `a' indicates a cluster of points that are discussed in the text.}
\label{fig:dp_minor_axis}
\end{figure*}

\begin{figure*}
\includegraphics[width=0.8\textwidth]{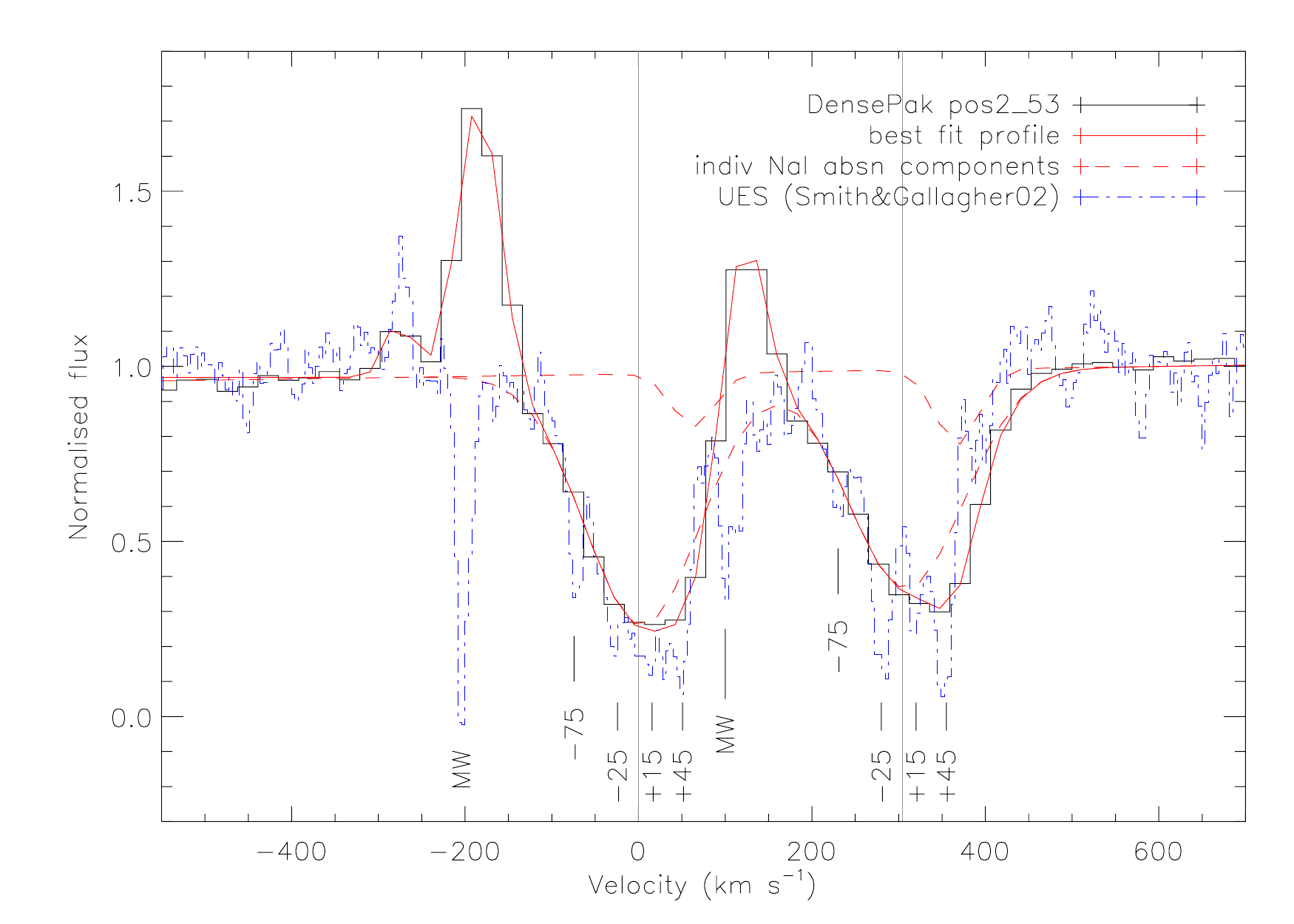}
\caption{A comparison of the Na\one\ profile of M82-F, a super star cluster to the west of the nuclear regions (see Fig.~\ref{fig:HSTfinder}), from our DensePak data (resolution 45~\kms) and that measured by \citet{smith_gall01} from their WHT/UES spectrum (resolution 8~\kms). The velocity scale is relative to the blue component of the Na\one\ doublet, and is shown relative to the systemic velocity of M82. The best fit profile and the individual components identified in the DensePak spectrum are shown in red. The vertical lines indicate the wavelengths corresponding to the systemic velocity for each of the Na\one\ lines. Absorption components identified in the UES spectrum by \citet{smith_gall01} are marked, together with absorption from the Milky Way (MW). In the DensePak spectrum, this Galactic absorption is filled by the telluric emission.}
\label{fig:m82-f_comp}
\end{figure*}

By reanalysing the CO data from \citet{walter02} and performing a detailed multi-component line fit, we find conclusively that molecular CO is entrained in the wind in these inner regions. In the southern wind plume (Fig.~\ref{fig:dp_minor_axis}), we identify two components of CO emission, equivalent to the line splitting seen in \citet{walter02}. The strongest component traces the redshifted (far-side) wall of the outflow cone, however a weaker component (C2) also traces the blueshifted side in some places (as seen in the pseudoslit 4 p-v plot). These results represent clear evidence that the inner wind contains entrained molecular gas that is coupled kinematically to the ionized gas phase.

A region of blueshifted ($-100$~\kms) CO emission can be seen in the south of the C2 velocity map (Fig.~\ref{fig:vel_CO}) at coordinates around ($-30$,$-10$), or in the pseudoslit 4 p-v plot (Fig.~\ref{fig:dp_minor_axis}, right panel) between offsets of $-20$ and $-35$. By comparing this to the minor axis p-v diagram shown by \citet{walter02}, we see that this component is equivalent to the faint blueshifted ``prong'' of the p-v fork on the southern side (their fig.~4). We therefore associate this component with the wind. Since this ``prong'' is only seen at low surface brightness, we do not identify it as clearly here since we masked low S/N data in the CO cube (see Section~\ref{sect:CO_obs}). It is interesting to note that this component is not seen directly along the minor axis, but only to the east.

On the opposite (north) side of the disk, we also find redshifted (30--100~\kms) CO emission in both C1 and C2 (seen in both minor axis pseudoslits). These velocities are partly consistent with the ionized gas measurements (particularly those from \citetalias{westm09a}), although the measurements of \citet{mckeith95} and \citet{shopbell98} that extend further out indicate a much faster acceleration in the cone back-wall than is seen in CO \citep[see fig.~4 of][]{walter02}. The fact that these CO components stay at very constant velocity with increasing disk height implies that they may trace more static clouds in the inner halo that could be pre-existing or tidal in origin. 


\citet{wills02} found evidence for four expanding shells in the inner disk of M82 from their high-resolution VLA A-configuration H\one\ data. The CO velocity splitting seen on the eastern side of the disk (see the major axis p-v plots in Fig.~\ref{fig:dp_major_axis}) is consistent with that measured by \citet{wills02} for their shells 1 and 2. On the western side, however, CO counterparts to the H\one\ shells 3 and 4 cannot be identified, although CO line splitting consistent with the known molecular superbubble $\sim$10$''$ west of the nucleus \citep{weis99} is clear.


\subsection{A counter-rotating Na\one\ component} \label{sect:counter_rot}

Throughout the central regions (IFU positions 1,2,4), we see what appears to be a counter-rotating neutral gas component in the form of Na\one\ C2 (see profile+fit examples Fig.~\ref{fig:eg_fits}b and c). These Na\one\ C2 velocities can clearly be seen in the major axis p-v diagram (Fig.~\ref{fig:dp_major_axis}), and do not match anything seen in the ionized or molecular gas. Furthermore, the extent of these apparently counter-rotating Na\one\ C2 velocities do not correspond to anything seen in optical imaging, including the dust lanes seen in the \textit{HST} I-band image (Fig.~\ref{fig:HSTfinder}), or the CO map.

M82-F is located at a projected distance of $\sim$$30''$ to the west of the nucleus and slightly above mid plane, just within the region where we find these unusual Na\one\ velocities. As mentioned above, \citet{smith_gall01} observed M82-F at high spectral resolution (R=45000; 8~\kms) and found a number of discrete narrow components in the Na\one\ profile. Two of them are redshifted at 15 and 45~\kms, and these correspond well with the C2 component identified in our DensePak spectra, which in this location is redshifted by 20--40~\kms\ (as shown in Fig.~\ref{fig:m82-f_comp}).

\citet{wills00} presented $1\farcs4\times1\farcs2$ spatial resolution VLA observations of M82 with a velocity resolution of $\sim$10~\kms. For the few SN remnants (SNRs) that were clearly separated from other nearby sources, they were able to detect multiple H\one\ absorption components and measure their velocities. SNR 39.10+57.3, located $\sim$~5$''$ west of the nucleus, 5$''$ below the mid-plane, was found to have four velocity components at +60, +35, $-10$, and $-120$~\kms\ (relative to systemic). These redshifted components are consistent with our Na\one\ C2 velocities. The H\one\ profiles against the another three SNR discussed in their paper located $\sim$25$''$ to the west of nucleus (41.95+57.5), and $\sim$20$''$ and $\sim$25$''$ east of the nucleus (45.17+61.2 and 6.70+67.1), however, did not exhibit any absorption components outside the allowed values of disk rotation.

What could this counter-rotating component represent? \citet{smith_gall01} comment that ``this gas could be associated with infalling tidal debris or gas outflow.'' Although the outflow interpretation is difficult to reconcile with the other known wind tracers, the possibility that this component represents gas inflow is very interesting. Some simulations of galactic-shock accretion of gas in an oval potential, such as a nuclear bar, have produced highly eccentric streaming motions toward the nucleus, some portion of which are counter-rotating \citep{wada98}. 

The H\one\ gas filaments in the halo of M82 are found in the velocity range of $-70$ to 70~\kms\ relative to systemic \citep*{yun_ho_lo93}.  Some of the Na\one\ features could, therefore, originate in extraplanar gas. In particular the anomalous Na\one\ velocities on the western side of M82 could represent an extension of the north-western H\one\ filament across M82 \citep{yun_ho_lo93}.  As the south-western H\one\ filament appears to join M82 at velocities close to the rotation speed, it is difficult to unambiguously distinguish between absorption from this filament and the outer disk of M82.  The possibility remains, however, that the south-western H\one\ filament is feeding gas into M82.

Nevertheless, the edge on inclination and the superposition of disk and bar orbits with complex projection effects make it impossible to be sure from these data whether this counter-rotating component is indeed a sign of gas inflow or not. Higher sensitivity H\one\ imaging at good spatial and velocity resolution of the central few kpc of the disk would be needed to investigate this further.



\subsection{Consequences for interpreting Na\one\ outflows vs.\ ionized gas}

Gas outflows have been identified in many star forming galaxies from low to high redshift \citep{heckman90, pettini02, tremonti07, weiner09}. Because of observational constraints it is common to measure a single outflow speed for a galaxy from either an integrated spectrum or a single long-slit pointing. Typically these are derived from measurements of interstellar absorption lines \citep[such as Na\one, e.g.][]{rupke05a, chen10}, but it is also common to use ionized emission lines \citep[e.g.][]{forster09}. Very few studies discuss the implications of using neutral vs.\ ionized and absorption vs.\ emission lines on determining the presence of galaxy outflows and their velocities.

This study highlights the inherent complexities of the relationship between the ionized and neutral phases, and therefore the assumptions of similarity between these phases and techniques. M82 is one of the nearest and best studied examples of a starburst and superwind system, and one that is often used as a template for more distant objects that cannot be resolved as well. However, spatially resolved Na\one\ and H$\alpha$ kinematics show that they cannot be used as equivalent tracers of the same gas. In H$\alpha$, a double-peaked line profile characteristic of the expanding superwind is clearly detected 10--20$''$ (few 100~pc) above and below the midplane \citep[][\citetalias{westm09a}]{mckeith95}. In Na\one, however, we find no clear evidence of an equivalent outflow component. 

\begin{figure}
\includegraphics[width=0.49\textwidth]{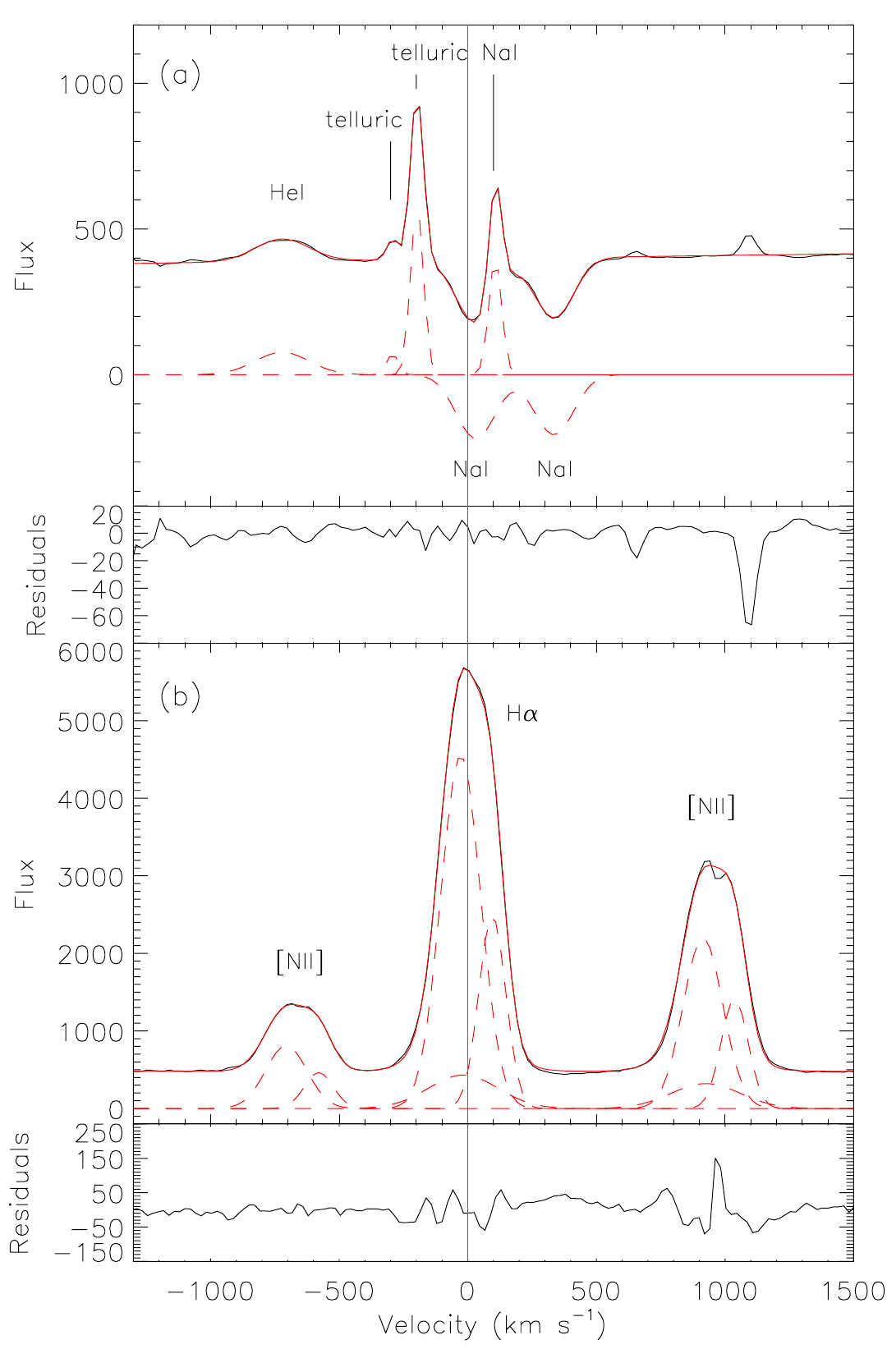}
\caption{The summed Na\one\ and H$\alpha$-[N\two]} spectra from DensePak IFU positions 1, 2 and 4 (i.e.\ the central $\sim$1\,$\times$\,1~kpc of the disk), shown together with the best fit to the emission and absorption lines, the individual components to the fit, and the residuals. For Na\one\, only one absorption component can be identified with a velocity centroid of +23~\kms\ (relative to systemic) and FWHM = 174~\kms. In contrast, three components can be identified in the H$\alpha$ and [N\two] lines.
\label{fig:int_spec}
\end{figure}

In Fig.~\ref{fig:int_spec} we show the summed Na\one\ and H$\alpha$-[N\two] profiles over DensePak IFU positions 1, 2 and 4 (i.e.\ the central $\sim$1\,$\times$\,1~kpc of the disk), together with our best fitting multi-Gaussian fit and the fit residuals. For Na\one, only one absorption component can be identified, with a velocity centroid of +23~\kms\ (relative to systemic) and FWHM = 174~\kms\ (mostly reflecting broadening due to rotation). Since no blueshifted component is seen, from this spectrum alone one would conclude that M82 does not host an outflow. In contrast, three line components can be identified in the H$\alpha$ and [N\two] lines, including two separated narrow components characteristic of outflowing material, and a broad (FWHM = 320~\kms) underlying component.

We have shown that at higher resolution the absorption profile can break up into many narrow components (Section~\ref{sect:neutral}), and it may be that the summed profile would do the same. However, not all studies looking for absorption line outflows are done at high resolution (for example the DEEP2 spectra used by \citealt{weiner09} and \citealt{kornei12} were taken at R=5000, which is similar to this study). Even then, the lines-of-sight that can be probed by absorption line studies rely on a bright background source, and this may not correspond with a region in which an outflow can be detected. For example, the high-resolution pencil-beam studies of \citet{smith_gall01} and \citet{schwartz04} did not cover regions in M82 where the H$\alpha$ outflow is most obvious \citep{shopbell98, mckeith95, westm09a}.

In conclusion, the high spectral resolution (R$\gtrsim$10000) observations are clearly preferable in order to search for signatures of a wind, so that the Na\one\ line profile is sampled at a sufficient level. We have shown that high velocity components are hard to find in highly inclined systems, and may easily be missed in observations of similar systems at a similar spectral resolution to ours.


\section{Summary} \label{sect:summary}

We present spatially resolved (3$''$ resolution) observations of the interstellar Na\one\ D $\lambda\lambda$5890,5896 doublet absorption obtained with the DensePak IFU on the WIYN telescope, and compare them to the $^{12}$CO (1$\rightarrow$0) (2.6~mm, 115~GHz) emission obtained with the OVRO millimetre array, across the inner $\sim$$2\times1$~kpc of the disk of M82. By measuring the Na\one, CO, and ionized gas (H$\alpha$) kinematics \citepalias[from][]{westm09a} at the same spatial resolution, and employing the same line fitting method, we have been able to make meaningful comparisons between the three gas phases. Our main conclusions are the following.

In general, the Na\one\ kinematics do not match those of H$\alpha$ or CO, although in the eastern part of the inner bar region the three tracers do match one another closely, indicating that we are seeing molecular, neutral and ionized gas all following the bar orbits.
The Na\one\ rotation curve is flatter both along the mid-plane and at disk heights of a few hundred pc. The likely explanation of this is that the Na\one\ line, being in absorption, probes gas that is located in front of the continuum source (the starburst), and is therefore biased towards material further out along the line-of-sight and thus with a lower radial component of its orbital velocity. Despite this, however, the rotation seen in Na\one\ C1 is consistent with same offset axis as that found from the radio and optical recombination lines \citep[][\citetalias{westm09a}]{r-r04} of PA=12$^{\circ}$.

We detect a component of the Na\one\ line with velocities that are forbidden by the known orbital rotation of the galactic disk. Hints of this component were seen in high spectral resolution Na\one\ observations of the cluster M82-F \citep{smith_gall01} and the H\one\ absorption profile towards a SNR to the south-west of the nucleus \citep{wills00}. We interpret this as material that is counter-rotating or perhaps inflowing towards the starburst. These velocities have not been found in any observation at any other gas temperature phase, suggesting that this material must be cold. To investigate this further, higher sensitivity H\one\ observations of the inner few kpc of the disk are needed.

We find a clear indication of entrained CO gas in the outflow with kinematics consistent with that of the H$\alpha$-emitting wind, thereby confirming the result of \citet{walter02}. It is possible that this CO line splitting follows the expanding shells detected in H\one\ absorption measurements by \citet{wills02}, and certainly traces the known molecular superbubble identified by \citep{weis99}. The existence of such structures might be expected to result from the massive and highly clumped young complexes in the M82 starburst zone. The breakout of hot gas from such high pressure bubbles would offer a likely source for the hot superwind.

However, we find no evidence that Na\one\ absorption is kinematically associated with the outflow. Since it would be rather surprising that the outflow does not entrain any neutral material (given that we know there is ionized and molecular gas in the flow), we conclude that a combination of (1) lack of velocity resolution and (2) confusion of the absorption profile due to the high inclination of the system is acting to prevent the detection. This view is supported by the observation of narrow, blueshifted Na\one\ components at higher resolution (R$>$30000) along select sight-lines towards M82 \citep{smith_gall01, schwartz04}.


Thus, in the search for neutral outflows from galaxies, high velocity components may easily be missed in observations at similar spectral resolution to ours. This is particularly true for highly inclined systems. Often, the inclination of a galaxy is not known (especially for more distant systems) so a non-detection of high velocity Na\one\ absorption can simply be an inclination effect. Higher spectral resolution (R$\gtrsim$10000) is always preferable when studying ISM absorption lines, and particularly so in these situations.


\section*{Acknowledgments}
The research leading to these results has received funding from the European Community's Seventh Framework Programme (/FP7/2007-2013/) under grant agreement No 229517. JSG gratefully acknowledges partial support of this research by grant NSF AST0708967 to the University of Wisconsin-Madison.

\bibliographystyle{mn2e}
\bibliography{/Users/mwestmoq/Dropbox/Work/references}
\bsp

\clearpage


\clearpage

\label{lastpage}
\end{document}